# Permittivity-asymmetric *quasi*-bound states in the continuum

Rodrigo Berté[1,2,*], Thomas Weber[1], Leonardo de S. Menezes[1,3], Lucca Kühner[1], Andreas Aigner[1], Martin Barkey[1], Fedja J. Wendisch[1], Yuri S. Kivshar[4], Andreas Tittl[1,*], Stefan A. Maier[5,1,6]

[1]Chair in Hybrid Nanosystems, Nano Institute Munich, Faculty of Physics, Ludwig-Maximilians-University Munich, Königinstrasse 10, 80539 München, Germany

[2]Instituto de Física, Universidade Federal de Goiás, 74001-970 Goiânia-GO, Brazil

[3]Departamento de Física, Universidade Federal de Pernambuco, 50670-901 Recife-PE, Brazil

[4]Nonlinear Physics Centre, Research School of Physics Australian National University, Canberra ACT 2601, Australia

[5]School of Physics and Astronomy, Monash University, Clayton Victoria 3800, Australia

[6]The Blackett Laboratory, Department of Physics, Imperial College London, London, SW7 2AZ, United Kingdom

*Corresponding authors: r.berte@physik.uni-muenchen.de , andreas.tittl@physik.uni-muenchen.de

**Abstract:**
**Broken symmetries lie at the heart of nontrivial physical phenomena[1,2]. Breaking the in-plane geometrical symmetry of optical systems allows to access a set of electromagnetic states termed symmetry-protected *quasi*-bound states in the continuum (qBICs)[3-6]. As optical systems are mainly defined by the geometry and the permittivity of its constituents, here we demonstrate, theoretically, numerically and experimentally, that such optical states can also be accessed in metasurfaces by breaking the in-plane symmetry in the permittivity of the comprising materials, showing a remarkable equivalence to their geometrically-asymmetric counterparts. However, while the physical size of atoms imposes a limit on the lowest achievable geometrical asymmetry, weak permittivity modulations due to carrier doping and electro-optical Pockels and Kerr effects, usually considered insignificant, open up the possibility of infinitesimal permittivity asymmetries for on-demand, and dynamically tuneable optical resonances of extremely high quality factors. We experimentally probe the excitation of permittivity-asymmetric qBICs ($\varepsilon$-qBICs) in the near infrared using a prototype Si/TiO$_2$ metasurface, in which the asymmetry in the unit cell is provided by the refractive index contrast of the dissimilar materials, surpassing any unwanted asymmetries from nanofabrication defects or angular deviations of light from normal incidence. $\varepsilon$-qBICs can also be excited in 1D gratings, where quality-factor enhancement and tailored interference phenomena via the interplay of geometrical and permittivity asymmetries are numerically demonstrated. The emergence of $\varepsilon$-qBICs in systems with broken symmetries in their permittivity may enable to test time-energy uncertainties in quantum mechanics[7], and lead to a whole new class of low-footprint optical and optoelectronic devices, from**

**arbitrarily narrow filters[8] and topological sources[9,10] , biosensing[11] and ultrastrong light-matter interaction[12] platforms, to tuneable optical switches[13].**

**Main Text:**
Symmetries, and the lack thereof, are deeply embedded in nature's fundamental and emergent phenomena[1,2]. The concept of symmetries underpins our subjective notion of aesthetics[14] as well as fundamental conservation laws, where conserved quantities are uniquely related to a preserved symmetry[15]. Broken symmetries are, however, no less interesting. Their effects manifest in phenomena as varied as the ordered but aperiodic arrangement of atoms in quasicrystals[16], parity (P) violation in nuclear beta decay[17,18], the preferred chirality of amino acids and sugars in biological molecules[19], the alignment of magnetic dipoles in a given direction below the Curie temperature in ferromagnets[20], among others[21-23]. The universality of broken-symmetry concepts is not just related to basic natural phenomena, but has also led to them being explored in a wide range of applications, from the synthesis of homochiral compounds[24] to the asymmetric flow of electric current in p-n junctions[25]. In optics, anti-symmetric profiles of the imaginary part of the refractive index, exploiting the interplay of gain and loss in coupled systems that are symmetric under parity and time transformations (PT-symmetric optical systems)[26], have allowed unidirectional light propagation[27,28], and also counterintuitive effects such as loss-induced transparency[29], and loss-induced lasing[30].

Similarly to PT-symmetric systems, the concept of bound states in the continuum (BICs) also stems from quantum mechanics[6,31]. First developed by Von Neumann and Wigner, who proposed[32] that an engineered potential could lead to the spatial localization of an electronic wave function of positive energy, BICs were later recognized to be an ubiquitous wave phenomenon. Accordingly, their effects have been demonstrated in optical[5], elastic[33], acoustic[34] and hydrodynamic[35] systems. Different kinds of BICs, essentially defined by the mechanism of light localization through which they decouple from the radiation continuum, have been reported in optical systems: symmetry-protected[4,5], single-resonance parametric BICs[36], BICs derived from hopping rate engineering in coupled waveguide arrays[37], Fabry-Pérot BICs[38], and BICs due to mode-interference within the same cavity, known as Friedrich-Wintgen BICs[39,40], to name just a few[6]. The Friedrich-Wintgen BICs, being realized in finite-sized systems, have an upper limit in the achievable quality-factors, although a completely decoupled resonance - a true BIC - can be achieved in the extreme conditions of null values of either the permittivity of shell layers or the diameter of the dielectric core in composite spheres[41].

While true symmetry-protected BICs are theoretical entities of infinitely high quality factors that can be only realized in lossless systems, which are spatially infinite in at least one dimension[6], *quasi*-BICs (qBICs), also termed supercavity modes[42], can be excited in finite structures and are able to couple to radiation impinging from the far-field. These signatures can be probed through geometric perturbations in the symmetry of the unit cell - for instance, via changes in the length, relative angle or area of the constituent resonators[3,43], their relative displacement in superlattice metasurfaces[44], or angular deviations from normal incidence of light in photonic crystals (PhCs)[4]. Apart for extended 2D arrays, other symmetries have also been explored, such as in radially-distributed systems[45,46]. Along with polaritons mediated by plasmons[47] and phonons[48], qBICs have become one of the main tools to manipulate electromagnetic fields and light-matter interaction at the nanometric scale. Even though polariton-mediated resonances are unrivalled in terms of near-field enhancements, the quality factors provided by qBICs are exceptionally high and can be tailored by the degree of asymmetry in the system[3], not being merely a function of the optical properties of the materials bearing the excitations. Indeed, enhancement of nonlinear effects[43,49], sensing[11,50] and lasing[51] mediated by symmetry-protected qBICs have been reported.

Here, given the nontrivial emergence of *quasi*-BICs in systems with broken symmetry, we propose and demonstrate that symmetry-protected *quasi*-BICs can be induced in geometrically-symmetric metasurfaces by breaking the in-plane symmetry of the unit cell regarding the isotropic permittivity of its constituents. We demonstrate the inverse-square dependence of the quality factor ($Q$) with the permittivity asymmetry and the remarkable equivalence of the optical response to their geometrically-asymmetric counterparts, as proposed in Fig. 1. Although most materials are commonly plagued by their small electro-optical coefficients, where only minute variations of the refractive index are obtained through gating, here we propose that "less is more" and that a strong modulation of the optical response can be obtained by a small perturbation of the refractive index symmetry of the system, with exquisitely high quality factors. Thermal perturbations have been used to modulate the refractive index through thermo-optic effects and break the symmetry in coupled waveguides[5]. Our work generalizes the principle to permittivity-asymmetric metasurfaces, where we rigorously demonstrate the condition for the existence of permittivity-asymmetric qBICs ($\varepsilon$-qBICs) and the dependence of quality factors with the asymmetry parameter. In contrast to typical predefined resonances and excitations, which can only be spectrally shifted through a modulation in the refractive index, $\varepsilon$-qBICs originate from the symmetry modulation itself, increasing the degrees of freedom of traditional optical systems and allowing precise control of the linewidth and spectral position of resonances, and also to turn them on and off at will.

We employ the formalism developed for the characterization of the dyadic Green's function of periodic open systems and its expansion into the orthogonal eigenmodes of the photonic structure[3,52] and apply it to permittivity-asymmetric metasurfaces (see Supplementary Information for derivations: equations S1-S42 and Figures S1 and S2). Similarly to its geometrically-asymmetric counterpart[3], the transmittance $T$ of a permittivity-asymmetric metasurface can be written in the structure of a Fano formula, in which the Fano asymmetry parameter $q$ (eq. S13a) becomes ill-defined for a perfectly symmetric unit cell (Section 1, Supplementary Information), corresponding to a true BIC situation.

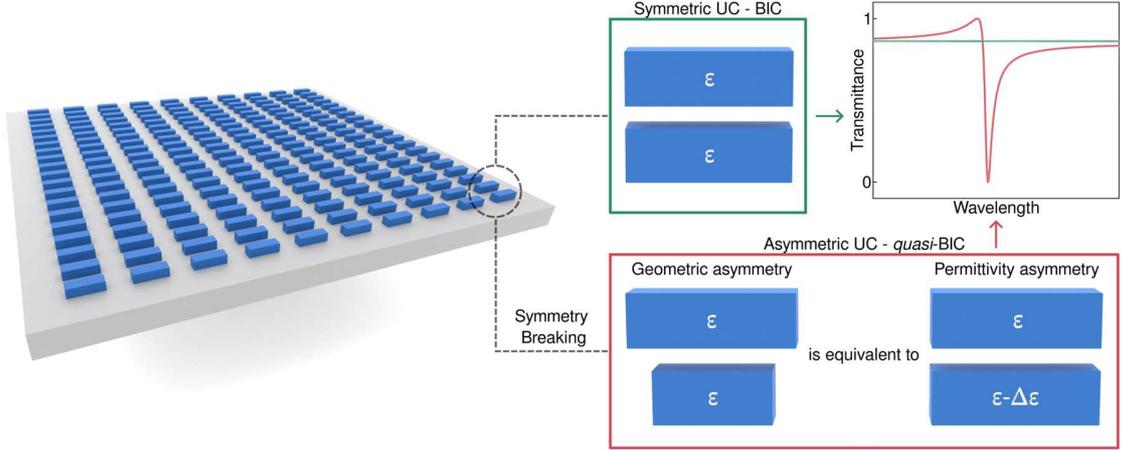

Figure 1: **BICs and *quasi*-BICs from the equivalence of asymmetric systems.** Proposed equivalence between unit cells (UCs) of geometrically- (bottom left) and permittivity-asymmetric (bottom right) metasurfaces (two-dimensional optical systems, in the left). The in-plane broken symmetries lead, in both cases, to the emergence of symmetry-protected *quasi*-BICs, which manifest as Fano lineshapes in the transmittance spectrum (in red, top right). Symmetric UCs correspond to a true BIC condition and the metasurface does not couple to an incident plane wave at normal incidence, resulting in a flat optical response (in green). The emergence of symmetry-protected *q*BICs in systems with broken symmetries cannot be intuitively, nor straightforwardly, derived from the optical properties of single resonators[1,2].

However, for a symmetry-protected BIC whose $x$-component of the electric field is odd (Fig. S2) with respect to an in-plane inversion transformation $((x, y) \to (-x, -y))$, i.e. $E_x(x, y) = -E_x(x, y)$, the condition for a true BIC can only be fulfilled if there is a symmetry in the permittivity relative to this pair of coordinates (eq. S16):

$$\varepsilon(k, x, y, z) = \varepsilon(k, -x, -y, z), \tag{1}$$

where $k = \omega/c$. Thus, the symmetry-protected BIC cannot be sustained in a system which is in-plane asymmetric in the permittivity. Or, that it only exists in the limit $\varepsilon(k, x, y, z) \to \varepsilon(k, -x, -y, z)$, where the Fano asymmetry parameter $q$ becomes ill-defined again. Indeed, numerical simulations (see Methods for details) show that breaking the in-plane symmetry in the permittivity of a metasurface containing a pair of geometrically identical resonators in the unit cell allows the excitation of *quasi*-BICs, as seen on the top panel of Fig. 2a. Because the symmetry in the permittivity is increasingly broken, the qBICs diverge spectrally from the true BIC state, shifting to longer or shorter wavelengths when the permittivity of the lower bar is increased ($+\Delta\varepsilon$, $\Delta\varepsilon > 0$) or decreased ($-\Delta\varepsilon$), respectively, relative to the unperturbed top bar. In addition, the linewidth of the resonances becomes wider for larger asymmetries in the permittivity, as quantified by the dimensionless asymmetry parameter $\alpha_\varepsilon = \Delta\varepsilon/\varepsilon$.

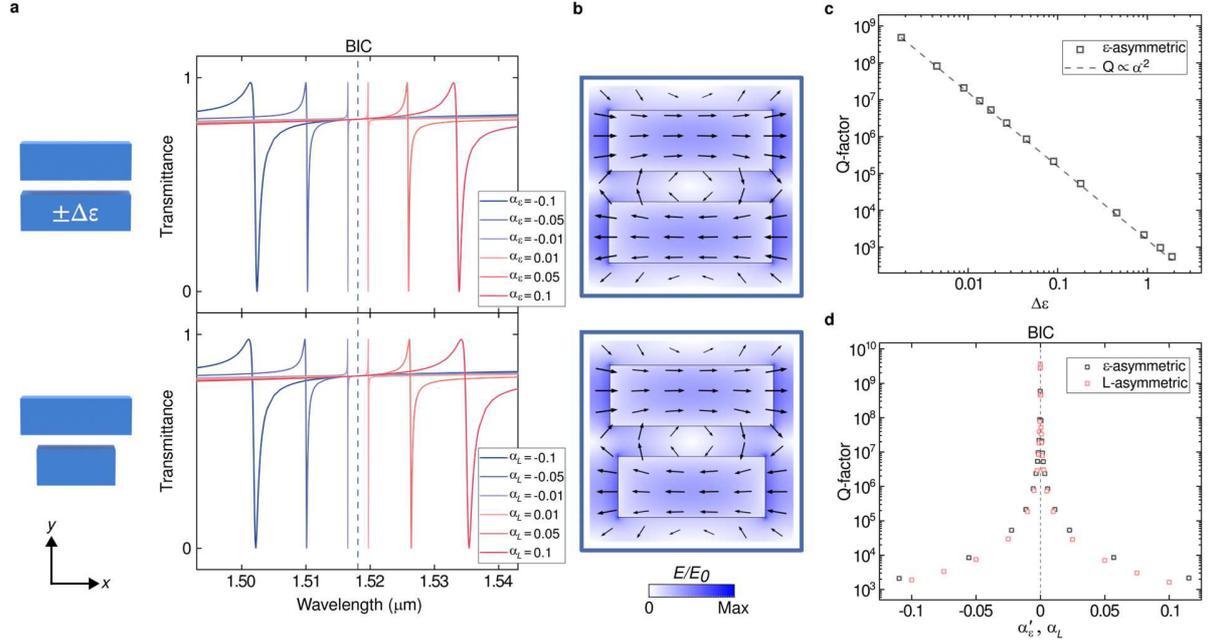

Figure 2: **Permittivity- and geometrically-asymmetric qBICs. a** Numerically calculated transmittance of permittivity- (top) and geometrically-asymmetric (bottom) metasurfaces. A dispersionless and lossless $n = \sqrt{\varepsilon} = 3$ material was used for the resonators of dimensions $0.8x0.3y0.1z$ μm (gap 150 nm and in-plane $xy$ pitch equal to 1 μm) on a SiO$_2$ substrate. The limiting case of the true BIC in a symmetric unit cell is shown as a vertical dashed line. **b** Near-field enhancement and polarization (arrows) of permittivity- ($\alpha_\varepsilon = -0.1$) and geometrically-asymmetric ($\alpha_L = -0.1$) qBICs. **c** Radiative quality factor ($Q_{rad}$, equivalent to the total quality factor $Q$ for a lossless material) as a function of the difference in permittivity ($\Delta\varepsilon$) between the resonators in the unit cell. Theoretically-predicted inverse-square dependence shown as a dashed line. **d** $Q_{rad}$ as a function of asymmetry parameters for permittivity- (black) and geometrically-asymmetric (L-asymmetric, red) metasurfaces.

It can be rigorously demonstrated that the radiative quality factor ($Q_{rad}$) of the permittivity-asymmetric system has an inverse-square dependence with the difference in permittivity between the resonators (Section 2, Supplementary Information, eq. S30):

$$Q_{rad} = \frac{2S_0}{k_0} \left| \int_{meta-atom\ \#2} d\boldsymbol{r}'\, E_{rs,x}(\boldsymbol{r}') \right|^{-2} |\Delta\varepsilon|^{-2} \qquad (2)$$

where $S_0$ is the area of the unit cell, $k_0$ the free-space wavenumber, $E_{rs,x}$ the $x$-component of the resonant state electric field, and the integration is performed in the region of the lower bar of perturbed permittivity (Figure S1, meta-atom #2). This theoretical prediction is in excellent agreement with numerical calculations, as seen in Fig 2c. When the geometric symmetry of the unit cell is broken instead, keeping the permittivity of the bars unperturbed, a nearly identical optical response is observed, regarding both the spectral shifts and linewidth of resonances for an equivalent geometrical asymmetry parameter ($\alpha_L = \Delta L/L_0$, being $L_0$ the original length of the bars). The qBICs shift to either longer or shorter wavelengths when the length of the lower bar is increased or decreased relative to the top bar, respectively. Since both permittivity- and geometrically-asymmetric unit cells share the same limiting symmetric case, the true BIC condition from which their qBICs diverge is also the same (vertical dashed line in Fig 2a). The similarity of these different qBICs is not limited to their far-field response, as the near-field enhancements and distributions are almost identical for equal asymmetry parameters (shown for $\alpha_\varepsilon = \alpha_L = -0.1$ in Fig 2b), even though the unit cells are substantially different from each other.

The equivalence of these two asymmetric systems can be further elucidated by considering the approximations of constant electric field magnitude within the asymmetry region $\Delta L$ and, as a coarser approximation, within the whole meta-atom #2, leading to an equality of radiative quality factors $Q_{rad}(\Delta\varepsilon) = Q_{rad}(\Delta L)$ (see section 3, Supplementary Information, eq. S41):

$$Q_{0_{\varepsilon,L}}|\alpha'_\varepsilon|^{-2} = Q_{0_{\varepsilon,L}}|\alpha_L|^{-2} \tag{3}$$

as long as the redefined asymmetry parameter $\alpha'_\varepsilon$ ($\alpha'_\varepsilon = \Delta\varepsilon/[\varepsilon(\omega_0) - 1]$, being $\omega_0$ the frequency of the qBIC) and $\alpha_L$ have identical values. We have defined a common coefficient $Q_{0_{\varepsilon,L}} = 2S_0/k_0 \left|E_{rs,x}\right|^{-2} |\sigma|^{-2}$ for both asymmetric systems, which denotes the fact that the different qBICs stem from the same symmetry-protected BIC. Here, $\sigma$ is the lateral cross-section of each meta-atom. In spite of the constant field approximations, numerical calculations show that the $Q_{rad}$ of both asymmetric systems behave almost identically with their respective asymmetries (Fig 2d).

Although the values of asymmetries used here are arbitrary, the choice of unit cell geometry imposes a clear limitation on how much the relative length of the bars can be changed and the geometrical asymmetries one may achieve. To exploit a parameter space of asymmetries of $\alpha_L < 10^{-4}$, for an 800nm-long resonator considered here, the length of one of the bars would need to be changed by $\Delta L = \alpha_L \times L_0 = 10^{-4} \times 800\,nm < 10^{-10}\,m$, the typical size of an atom, well below current state-of-the-art nanofabrication capabilities, making this parameter space virtually unattainable. Naturally, other geometrical asymmetries may be explored, such as the relative angle between resonators or the position of protrusions in the unit cell[43]. Nevertheless, all these geometries are fundamentally constrained by nanofabrication accuracy and, ultimately, limited by the atomic scale of possible spatial distributions of chemical elements, especially when pushing toward resonances in the near-IR and visible spectral ranges where characteristic dimensions become smaller. Minute changes in the refractive index, on the other hand, are easily achievable and routinely employed through carrier injection, chemical doping or electro-optical Pockels and Kerr effects[53], enabling access to this geometrically unavailable parameter space of asymmetries. Furthermore, the giant tuneability of the refractive index ($> 20\%$ near excitonic resonances) in transition metal dichalcogenides[54] allows the development of platforms where both small and large asymmetries can be accessed in a single structure.

Experimentally, we demonstrate the excitation of $\varepsilon$-qBICs using a carefully tailored metasurface geometry [see Methods for fabrication[55]] comprised of pairs of Si and TiO$_2$ resonators, where the asymmetry in the unit cell (Fig 3a) is provided by the permittivity contrast between the two materials ($\varepsilon_{Si} \sim 12$ and $\varepsilon_{TiO_2} \sim 4.9$ at 1.3 µm, see Fig S3 for ellipsometry data). The simplicity of this demonstration allows us to probe the emerging permittivity-driven resonances we have proposed beyond any asymmetries that might arise from imperfections in nanofabrication and off-normal angular dispersion of the incident light. These unwanted asymmetries would inevitably lead to $q$BICs and Fano-like features in the transmittance spectrum of geometrically-symmetric metasurfaces, complicating the interpretation of the spectral features. Instead, the large difference in permittivity between the materials generates spectrally wide resonance in transmittance, isolating the $\varepsilon$-qBIC behaviour and facilitating comparison with numerical models. Experimental transmittance spectra are shown in Fig 3a for different metasurface realizations where the in-plane dimensions in the unit cells were scaled to cover a range of resonance wavelengths, showing excellent agreement with their respective numerical simulations. Such wide resonances would require a significant geometrical asymmetry or a large incidence angle of light in geometrically-symmetrical systems to be realized, whereas in our samples geometrical deviations from nominal values come from imperfections in nanofabrication (with an estimated upper limit of a few nanometres) and experiments were performed using near-normal incident light, ruling out their contribution to the observed resonances.

$\varepsilon$-qBICs may also be explored in geometries with a higher degree of symmetry, such as in one-dimensional gratings (1DG, Fig. 4). Similar to the 2D metasurface, a symmetry breaking in the isotropic permittivity of adjacent ridges leads to the emergence of diverging symmetry-protected qBICs (Fig. 4b) of characteristic opposite polarization of the electric field within the unit cell (Fig. 4a). This simpler geometry facilitates sample biasing and the dynamic symmetry breaking of the permittivity through carrier injection or other electro-optical effects, easing the fabrication of active devices. For instance, by biasing both ridges simultaneously, increasing their permittivity while keeping the asymmetry $\alpha_\varepsilon$ constant, a continuous shifting of the $\varepsilon$-qBIC with nearly invariant linewidth can be achieved (Fig. 4c). Structures with higher symmetry also imply lower losses due to reduced surface roughness and additional edges present in resonators of 2D metasurfaces. A 1DG is then a suitable geometry when aiming for high and tunable Q-factors. Here, the same argument for atomically-limited asymmetries applies: for asymmetry parameters $\alpha < 10^{-4}$, the difference in width ($0.0325\ nm/325\ nm$), height ($0.01\ nm/100\ nm$) or relative displacement ($0.01\ nm/100\ nm$ gap) between adjacent ridges for the 1DG in Fig. 4a need all to be below 1 Å. Modulations in permittivity of 1 part in 10,000 are, on the other hand, easily achieved in silicon[53].

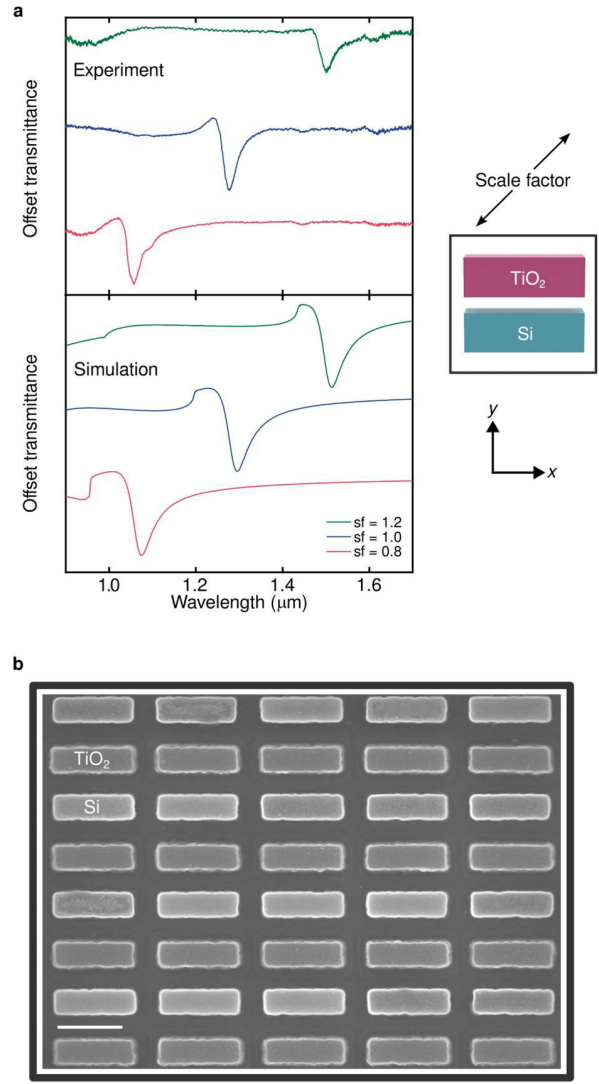

Figure 3: **Experimental demonstration of $\varepsilon$-qBICs. a** Permittivity-asymmetric qBICs in geometrically-symmetric metasurfaces. Although geometrically identical, the asymmetry in the unit cell that leads to qBICs is provided by the difference in permittivity between the resonators' materials (TiO$_2$ and Si). Experimental (top) and numerical (bottom) transmittance spectra of Si/TiO$_2$ metasurfaces of different in-plane ($xy$) scale factors ($sf$). Structures of dimensions $0.665x0.225y0.12z$ μm, gap $187.5\ nm$ and in-plane $825\ nm\ xy$ pitch were used for the unit cell ($sf = 1$, right-hand side scheme). Metasurfaces were fabricated on a SiO$_2$ substrate. Discrepancies are attributed to deviations from nominal fabrication values and inaccuracies in ellipsometry data (Fig. S3) fitting. **b** Scanning electron microscopy (SEM) image showing a few unit cells of the array. Both Si and TiO$_2$ structures can be clearly distinguished by their difference in contrast. Scale bar corresponds to 500nm.

Geometrical asymmetries, either deliberate or originating from imperfections in fabrication, might as well be compensated by asymmetries in the permittivity (Fig. 4d,e). For instance, given an arbitrary $10\ nm$ difference in height between adjacent ridges in the unit cell of the 1DG, which results in a geometrically-asymmetric qBIC (Fig. 4e, top), a further breaking in the permittivity symmetry leads to an enhancement in the quality factor of the pre-existing resonance ($> 6$-fold, in this particular configuration), albeit with a finite upper limit. This raises the question of whether a symmetric state - a true BIC - may exist in a system which combines asymmetries in its geometry and optical properties, for which the condition $D_{j_0,x}(k) = 0$ (eq. S6) is met. Interestingly, the Fano parameter ($q$) changes sign twice (Fig. S4) as the resonance shape flips horizontally and crosses the original BIC state, suggesting a change in the coupling strength between the discrete (the qBIC) and the continuum states[56], unlocking a way of engineering this key parameter and controlling interference processes on the nanoscale by combining different asymmetries.

In conclusion, we propose and demonstrate the excitation of symmetry-protected *quasi*-BICs in metasurfaces that are asymmetric regarding their permittivity in the unit cell (termed $\varepsilon$-qBICs), showing a remarkable equivalence to their geometrically-asymmetric counterparts, which, in a sense, may be interpreted as a localized perturbation in the permittivity of an otherwise symmetric system. Whereas the parameter space of possible asymmetries is fundamentally constrained by atomic length scales in geometrical symmetry breaking, permittivity-driven asymmetries do not exhibit a clear lower limit. For resonances in the visible and near IR range, asymmetries of $10^{-4}$ require structural changes that are smaller than 1 Å, while such modulations in the permittivity of CMOS-compatible materials have been demonstrated decades ago[53]. Going beyond our demonstrations for one- and two-dimensional arrays, we believe that the concept of $\varepsilon$-qBICs can be extended to 3D metamaterials and others of different dimensionality[56]. As $\varepsilon$-qBICs originate from the modulation of the permittivity itself, the paradigm of the spectral shifting of predefined resonances can be extended, implying the possibility of active control over their linewidths and existence.

Permittivity asymmetries could be used to compensate not only for pre-existing geometrical asymmetries, enhancing Q-factors and tailoring interference phenomena as shown before, but also for detrimental effects due to a nonlinear material response in higher harmonic generation[57]. Additionally, the topological nature of symmetry-protected BICs[58] as vortex centres of the far-field polarization combined with the manipulation of permittivity symmetries may impart further control on the generation and detection of optical vortices (light with orbital angular momentum) in planar devices[10] – a required geometry for future optoelectronic platforms[59]. Usually deemed robust against large structural variations, the ability of topological states to withstand symmetry perturbations (and keep their polarization vortex nature) needs to be properly assessed, as, in some cases, significant changes in the electric field profile from the symmetry-protected BIC to the symmetry-protected *quasi*-BIC state (which also lies at a high symmetry point in the reciprocal lattice) follow from the breaking of in-plane inversion symmetry[60].

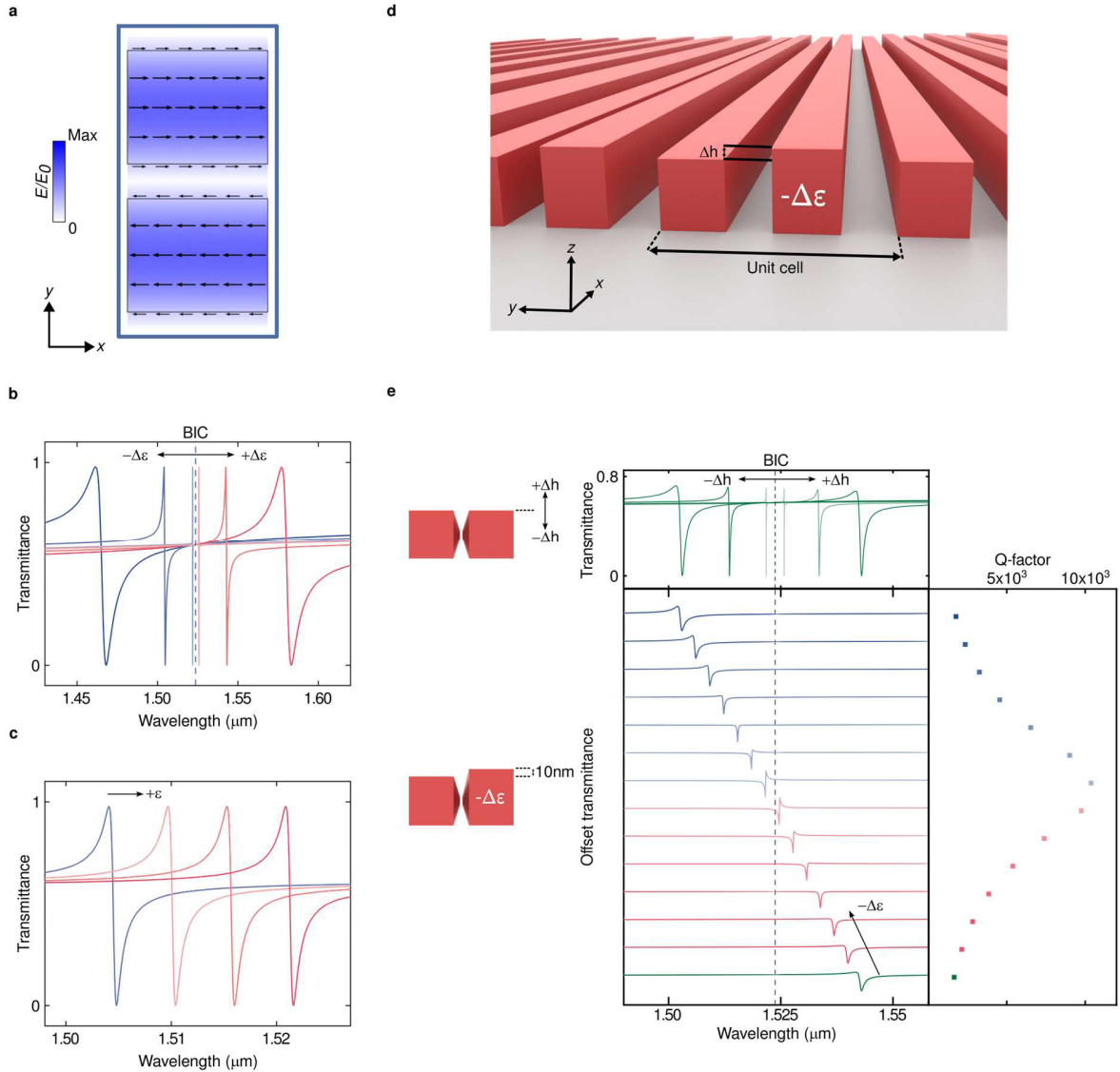

Figure 4: **ε-qBICs in a structure with a higher symmetry (1D grating) and compensation of geometrical asymmetries. a** Near-field enhancement and typical opposite polarization (arrows) of a symmetry-protected qBIC in a 1D grating of $100\,nm$ high, $325\,nm$ wide permittivity-asymmetric ridges, with a $100\,nm$ gap between ridges and a $850\,nm$ $y$-pitch in the unit cell. A $400\,nm$ wide section of the 1D grating along the $x$ axis is shown. A lossless $n = 3$ material was used for the ridges. **b** Isotropic permittivity of adjacent ridges vary by $\Delta\varepsilon$. Resonances diverge from the true BIC and widen as $\Delta\varepsilon$ magnitude increases. **c** Simultaneous bias and increase in permittivity ($+\varepsilon$) of adjacent bars allows qBIC shifting with nearly constant $Q$, as long as the permittivity asymmetry is kept constant. **d** Schematic representation of a 1D grating with simultaneous height ($\Delta h$) and permittivity ($\Delta\varepsilon$) asymmetries in the unit cell. **e** (top, in green) qBICs induced by height asymmetries ($\pm\Delta h$), diverging from the true BIC state (vertical dashed line) of identical ridges. (bottom) Geometric asymmetry compensation via permittivity symmetry breaking ($-\Delta\varepsilon$) in the unit cell. Starting from a $10\,nm$ asymmetry in height (bottom green curve), sequential

decrements of 1% in the permittivity in the taller ridge allow a > 6-fold enhancement in the Q-factor (right) and a tailoring of the Fano parameter $q$ as the resonance flips horizontally when crossing the BIC.

Platforms based on $\varepsilon$-qBICs may also allow to test the practical limits of how spectrally narrow a two-dimensional device might be, the assumption of arbitrary precision of observables in quantum mechanics and the validity of time-energy uncertainties[7]. For this realization, however, effects of permittivity self-modulation and momentum-space dispersion due to finite-sized structures and angular distribution of incoming radiation[9] need to be taken into account. When pushing for higher Q-factors, truly low-loss materials and small size dispersion of structures become relevant. Nanometre-size standard deviations (for structures of typical dimensions around a few hundred nm) have reduced numerically-predicted Q-factors of 2D metasurfaces - inherently more lossy than 1DG - from 20-50%, depending on the unit cell design[61]. This deleterious effect should be mitigated over time as electron beam lithography protocols are improved, with sub-nanometric standard deviations already reported in conventional resists[62], reaching atomic precise manufacturing using aberration-corrected STEM in bottom-up approaches[63] - through which, in principle, virtually defect-free structures can be fabricated. From a set of fundamentally-limited metasurfaces, one may envision a cavity-free generation of continuous-wave frequency combs in on-chip low-dimensional structures, with a vast number of applications in sensing and metrology[64]. Although requiring a deliberate symmetry breaking, the emergence of symmetry-protected *quasi*-BICs in optical systems with broken symmetries in the permittivity, which cannot be intuitively, nor straightforwardly, derived from the spectral response of isolated resonators[1,2], provides additional degrees of freedom for the control of optical resonances. Nonetheless, it is reasonable to speculate whether a strong anisotropy, such as in polarization - an intensively investigated aspect of the cosmic microwave background[65] - , might be elicited by a spatially-coherent but small fluctuation in the optical properties of a medium. $\varepsilon$-qBICs may allow the development of novel optical devices and for traditional ones, be they linear or nonlinear sources, filters and switches or detectors, to be reimagined.

**Methods**

**Numerical calculations.** Eigenfrequency and full-wave solutions of Maxwell's equations in the frequency domain were performed using the commercially available RF module of the finite element solver COMSOL Multiphysics. Port boundary conditions were employed for the excitation of the structures and transmittance calculations, and perfectly matched layers (PMLs) domains for the absorption of propagating waves. A dispersionless refractive index of $n = 1.45$ was employed for the $SiO_2$ substrate, while the dispersive complex permittivities of the amorphous silicon and $TiO_2$ films, as measured via ellipsometry, were used for the resonators.

**Sample fabrication.** Samples were fabricated in a two-step process[55]. First, a-Si was deposited onto fused silica substrates via PECVD. A double-layer resist (PMMA 495k and 950k) was spin-coated onto the sample, which was subsequently patterned using EBL (Raith ELine Plus). A Cr hardmask was deposited via electron-beam evaporation in an ultra-high vacuum chamber. The samples were then etched using RIE with chlorine-based chemistry. For the second step, a layer of a-TiO$_2$ was deposited via reactive magnetron sputtering using a Ti target in an Ar/O$_2$ atmosphere (Angstrom Nexdep), followed by EBL, deposition of a Cr hardmask and RIE using fluorine-based chemistry. Finally, the hardmask was removed in a wet-etching process using Cr etchant standard (Sigma-Aldrich).

**Optical measurements.** In order to detect the eps-qBIC resonances in the near infrared (NIR), a white light source (NKT Photonics SuperK Extreme, ~35 ps pulses, 3.54 GHz rep. rate, 0.26 mW/nm, average power around 1400 nm) was used for illuminating the sample in the wavelength range between 900 nm and 1700 nm. The Si/TiO$_2$ metasurface was excited by a linearly polarized light oriented parallel to the main axis ($x$) of the resonators. The excitation beam was concentrated on the sample by an objective (10x, 0.25 NA) and the transmitted light was collected by a second objective (60x, 0.70 NA), being sent to a CCD camera (Qimaging QICAM Fast 1394) for imaging. The metasurface could then be located and optimally positioned under the excitation beam, typically attenuated with an OD 3.0 neutral density filter (Thorlabs NE30A). The total field of view has a diameter of ~40 μm, and the desired region on the sample was selected by using an iris placed right after the sample. With the aid of a flip mirror, the transmitted light could be coupled into a multimode optical fiber connected to a NIR spectrometer (30 cm focal length, 100 l/mm 1600 nm blazed grating) equipped with an liquid N$_2$ cooled InGaAs camera (Princeton Instruments OMA V:1024-1.7) for recording the transmission spectra. A reference spectrum was taken for normalization (transmittance) from a sample region nearby the metasurfaces (glass) before measuring the region of interest. Under typical excitation conditions, the integration times were 600 ms for acquiring the reference and transmission spectra.

## Acknowledgements

R.B. acklowledges the National Council for Scientific and Technological Development (CNPq, PDJ 2019 – 150393/2020-2). We acknowledge financial support from the Deutsche Forschungsgemeinschaft (DFG, German Research Foundation) under Grant Nos. EXC 2089/1−390776260 (Germanýs Excellence Strategy and TI 1063/1 (Emmy Noether Program), the Bavarian State Ministry of Science, Research, and Arts through the program "Solar Technologies Go Hybrid (SolTech)". Y.K. was supported by the Australian Research Council under grants DP210101292, and by the Strategic Fund of the Australian National University. S.A.Maier additionally acknowledges the Lee-Lucas Chair in Physics as well as the Engineering and Physical Sciences Research Council (EPSRC, EP/W017075/1) and the Australian Research Council. We thank local research clusters and centers such as the Center of Nanoscience (CeNS) for providing communicative networking structures.

## Data availability

The data that support the findings of this study are available from the corresponding authors upon reasonable request.

## Contributions

R.B conceived the project and performed the numerical analysis. R.B and Y.K performed the theoretical analysis. R.B, L.K. and A.T designed the experimental demonstration. T.W and L.K. fabricated the samples. L. de S.M., A.A., M.B. and F.J.W. performed the optical measurements. R.B., A.T. and S.A.M. supervised the project. R.B. wrote the manuscript. All other authors contributed to the manuscript improvement.

## Competing interest

The authors declare no competing interests.

## Additional information

Correspondence and requests for materials should be addressed to R.B. and A.T.

# Supplementary Information for Permittivity-asymmetric *quasi*-bound states in the continuum

Rodrigo Berté[1,2,*], Thomas Weber[1], Leonardo de S. Menezes[1,3], Lucca Kühner[1], Andreas Aigner[1], Martin Barkey[1], Fedja J. Wendisch[1], Yuri S. Kivshar[4], Andreas Tittl[1,*], Stefan A. Maier[5,1,6]

[1]Chair in Hybrid Nanosystems, Nano Institute Munich, Faculty of Physics, Ludwig-Maximilians-University Munich, Königinstrasse 10, 80539 München, Germany

[2]Instituto de Física, Universidade Federal de Goiás, 74001-970 Goiânia-GO, Brazil

[3]Departamento de Física, Universidade Federal de Pernambuco, 50670-901 Recife-PE, Brazil

[4]Nonlinear Physics Centre, Research School of Physics Australian National University, Canberra ACT 2601, Australia

[5]School of Physics and Astronomy, Monash University, Clayton Victoria 3800, Australia

[6]The Blackett Laboratory, Department of Physics, Imperial College London, London, SW7 2AZ, United Kingdom

*Corresponding authors: r.berte@physik.uni-muenchen.de , andreas.tittl@physik.uni-muenchen.de

## 1 - Radiation transmittance through a permittivity-asymmetric metasurface and the origin of $\varepsilon$-qBICs

A derivation essentially identical to the one performed by Koshelev et al[1] can be applied to obtain the radiation transmittance $T$ through a permittivity-asymmetric metasurface. Here we highlight the few differences between them. In our case, the meta-atoms of the unit cell (Figure S1) are pairs of geometrically-identical bars embedded in an isotropic medium of permittivity 1. Although geometrically identical, the permittivity $\varepsilon(\omega, r)$ of the bars varies isotropically by the dispersionless amount $\Delta\varepsilon$. It can be shown that the complex transmission amplitude $t$ of an $x$-polarized incident plane wave ($e_x E_0 e^{ikz} e^{-i\omega}$ ) through the metasurface, being $T = |t|^2$, has the form:

$$t = 1 + \frac{ik}{2S_0}\int d\mathbf{r}'[\varepsilon(k, \mathbf{r}') - 1]$$

$$-\frac{ik^2}{4S_0}\sum_j \frac{\left(\int d\mathbf{r}'[\varepsilon(k, \mathbf{r}') - 1]E_{j,x}(\mathbf{r}')e^{-ikz'}\right)\left(\int d\mathbf{r}''[\varepsilon(k, \mathbf{r}'') - 1]E_{j,x}(\mathbf{r}'')e^{ikz''}\right)}{k - k_j} \quad (S1)$$

$$-\frac{ik^3}{2S_0}\int d\mathbf{r}'[\varepsilon(k, \mathbf{r}') - 1]e^{-ikz'}\int d\mathbf{r}''[\varepsilon(k, \mathbf{r}'') - 1][\Delta\hat{G}(k, \mathbf{r}', \mathbf{r}'')]_{xx}e^{ik\ ''}$$

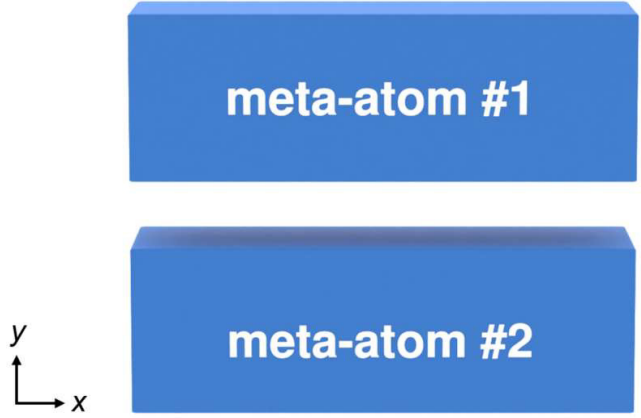

Figure S1: Unit cell of metasurface. Permittivity of meta-atom #2 varies isotropically by $\Delta\varepsilon$ relative to meta-atom #1.

where $k = \omega/c$, $S_0$ is the area of the periodic unit cell; $j$ is the index, $E_{j,x}$ the $x$-axis projection of the electric field and $k_j = \omega/c$ the spatial frequency of the resonant state, respectively. $[\Delta\hat{G}]_{xx}$ represents the $x$-axis projection of the contribution of Rayleigh anomalies, related to discontinuities in the dyadic Green's Function (GF) of the photonic structure $\hat{G}$.[2] The integral on the second term can be divided into three regions of the unit cell, inside each meta-atom and outside in the surrounding medium:

$$\frac{ik}{2S_0}\left(\int_{meta-atom\ \#1} d\mathbf{r}'[\varepsilon(k,\mathbf{r}') - 1] + \int_{meta-atom\ \#2} d\mathbf{r}'[\varepsilon(k,\mathbf{r}') + \Delta\varepsilon - 1]\right.$$
$$\left. + \int_{medium} d\mathbf{r}'[\varepsilon(k,\mathbf{r}') - 1]\right) \quad (S2)$$

which is non-zero only inside the meta-atoms where $\varepsilon(k,\mathbf{r}') \neq 1$. For an isotropic metasurface $\varepsilon(k,\mathbf{r}') = \varepsilon(k)$, the integrals are independent of the integrand and $\int d\mathbf{r}'$ results in the volume of each meta-atom, defined as $V_0/2$, then:

$$\frac{ik}{2S_0}\left(\int_{meta-atom\ \#1} d\mathbf{r}'[\varepsilon(k,\mathbf{r}') - 1] + \int_{meta-atom\ \#2} d\mathbf{r}'[\varepsilon(k,\mathbf{r}') + \Delta\varepsilon - 1]\right)$$
$$= \frac{ik}{2S_0}[\varepsilon(k) - 1]V_0 + \frac{ik}{4S_0}[\Delta\varepsilon]V_0 \quad (S3)$$

It results for $t$:

$$t(k) = 1 + \Delta t(k) + \frac{ik}{2S_0}[\varepsilon(k) - 1]V_0 + \frac{ik}{4S_0}[\Delta\varepsilon]V_0$$
$$- \frac{ik^2}{4S_0}\sum_j \frac{\left(\int d\mathbf{r}'[\varepsilon(k,\mathbf{r}') - 1]E_{j,x}(\mathbf{r}')e^{-ikz'}\right)\left(\int d\mathbf{r}''[\varepsilon(k,\mathbf{r}'') - 1]E_{j,x}(\mathbf{r}'')e^{ikz''}\right)}{k - k_j} \quad (S4)$$

where $\Delta t(k)$ is defined as:

$$\Delta t(k) = -\frac{ik^3}{2S_0}\int d\mathbf{r}'[\varepsilon(k,\mathbf{r}') - 1]e^{-ikz'}\int d\mathbf{r}''[\varepsilon(k,\mathbf{r}'') - 1][\Delta\hat{G}(k,\mathbf{r}',\mathbf{r}'')]_{xx}e^{ikz''} \quad (S5)$$

The coupling amplitude of the resonant state $E_{j,x}$ may be defined:

$$D_{j,x}(k) = \frac{k}{\sqrt{2S_0}}\int d\mathbf{r}'[\varepsilon(k,\mathbf{r}') - 1]E_{j,x}(\mathbf{r}')e^{ikz'} \quad (S6)$$

The geometrical symmetry of the system imposes that each resonant state is either symmetric or anti-symmetric relative to the z-axis. They can be classified with respect to the up-down mirror symmetry as $E_{j,x}(-z) = (-1)^p E_{j,x}(z)$, where $p = 0, 1$ for even and odd modes, respectively. The asymmetry in permittivity does not break the up-down mirror symmetry of the system, so we may keep this assumption, which implies that:

$$E_{j,x}(\mathbf{r}')e^{ikz'} = (-1)^p E_{j,x}(\mathbf{r}')e^{-ikz'} \quad (S7)$$

making both integrals on the last term of equation (S4) identical, apart from the term $(-1)^p$. Using the definition of the coupling amplitude, results for $t$:

$$t(k) = 1 + \Delta t(k) + \frac{ik}{2S_0}[\varepsilon(k) - 1]V_0 + \frac{ik}{4S_0}[\Delta\varepsilon]V_0 - i\sum_j \frac{(-1)^p[D_{j,x}(k)]^2}{2(k - k_j)} \quad (S8)$$

Although all modes $j$ contribute to the far-field scattered field, we may focus in the vicinity of the *quasi*-BIC, the resonant state of the photonic structure of index $j_0$ and parity $p_0$. This resonant state has eigenfrequency $ck_{j_0} = \omega_0 - i\gamma/2$, where $\omega_0$ is the frequency and $\gamma$ is the mode inverse radiation lifetime. Splitting the sum on the last term of equation (S8):

$$\sum_j \frac{(-1)^p[D_{j,x}(k)]^2}{2(k - k_j)} = \frac{(-1)^{p_0}[D_{j_0,x}(k)]^2}{2(k - k_{j_0})} + \sum_{j \neq j_0} \frac{(-1)^p[D_{j,x}(k)]^2}{2(k - k_j)}$$

$$= \frac{(-1)^{p_0}[D_{j_0,x}(k)]^2}{\frac{\gamma}{c}\left(\frac{2(\omega - \omega_0)}{\gamma} + i\right)} + \sum_{j \neq j_0} \frac{(-1)^p[D_{j,x}(k)]^2}{2(k - k_j)} \quad (S9)$$

By defining the parameters:

$$A(k) = \frac{(-1)^{p_0}[D_{j_0,x}(k)]^2}{\gamma/c} \quad (S10a)$$

$$B(k) = 1 + \Delta t(k) + \frac{ik}{2S_0}[\varepsilon(k) - 1]V_0 + \frac{ik}{4S_0}[\Delta\varepsilon]V_0 - i\sum_{j \neq j_0} \frac{(-1)^p[D_{j,x}(k)]^2}{2(k - k_j)} \quad (S10b)$$

$$\Omega = \frac{2(\omega - \omega_0)}{\gamma} \quad (S10c)$$

where we have chosen to incorporate the extra term $ik/4S_0 [\Delta\varepsilon]V_0$ due to the permittivity asymmetry into $B(k)$. We may write $t(k)$ then as:

$$t(k) = B(k) - i\frac{A(k)}{\Omega + i} \tag{S11}$$

from which the transmittance $T = |t|^2$ can be obtained. Although explicit expressions of $A(k)$, $B(k)$ and $\Omega$ should provide the asymmetric line shape for the transmittance expected for a *quasi*-BIC, we may rearrange $T$ explicitly in the structure of a Fano formula by treating $A(k)$, $B(k)$ and $\Omega$ as terms akin to free parameters instead. By using $A$ and $B$ in their polar form ($A = |A|e^{i\theta_A}$, $B = |B|e^{i\theta_B}$), through a lengthy algebra, $T$ can be finally written as:

$$T(\omega) = \frac{T_0(\omega)}{1+q^2(\omega)} \frac{[q(\omega) + \Omega]^2}{\Omega^2 + 1} + T_{bg}(\omega) \tag{S12}$$

where the new explicit and implicit parameters $T_0$, $T_{bg}$, $q$, $\psi$ and $\Delta$ are:

$$q(\omega) = \tan\psi \tag{S13a}$$

$$\psi(\omega) = \frac{1}{2}\cot^{-1}\left\{\cot\Delta - \frac{|A|}{2|B|\sin\Delta}\right\} \tag{S13b}$$

$$\Delta(\omega) = \arg(AB^*) \tag{S13c}$$

$$T_0(\omega) = |A||A - 2B| = |A|(|A|^2 - 4|A||B|\cos\Delta + 4|B|^2)^{1/2} \tag{S13d}$$

$$T_{bg}(\omega) = |B|^2 - \frac{T_0}{1+q^2} \tag{S13e}$$

Even though equation (S12) has the structure of a Fano formula, the asymmetric Fano lineshape only holds if the dependence of the terms $q(\omega)$, $\Delta(\omega)$, $T_0(\omega)$ and $T_{bg}(\omega)$ with the frequency $\omega$ is smooth around the frequency $\omega_0$ of the *quasi*-BIC, allowing it to be neglected. This is a reasonable assumption as long as neighbouring resonances $E_j$ ($j \neq j_0$) do not overlap with the *quasi*-BIC.

For symmetric metasurfaces, corresponding to a true BIC condition, equation (S12) should return a transmittance identical to that of the background $T_{bg}$ term, where only contributions of modes other than the *quasi*-BIC ($j \neq j_0$) are present. When dealing with a particular set of BICs whose field pattern is such that the $x$-component of the electric field is odd with respect to an in-plane inversion symmetry (($x, y$) → ($-x, -y$)):

$$E_{j_0,x}(x, y) = -E_{j_0,x}(-x, -y) \tag{S14}$$

this condition implies that the contribution of pairs of in-plane inverted coordinates in the integrand of the coupling amplitude $D_{j_0,x}(k)$ (equation (S6)) cancel out:

$$[\varepsilon(k,x',y',z') - 1]E_{j_0,x}(x',y',z')e^{ikz'}$$
$$+ [\varepsilon(k,-x',-y',z') - 1]E_{j_0,x}(-x',-y',z')e^{ikz'}$$
$$= [\varepsilon(k,x',y',z') - 1]E_{j_0,x}(x',y',z')e^{ikz'}$$
$$- [\varepsilon(k,-x',-y',z') - 1]E_{j_0,x}(x',y',z')e^{ikz'} = 0$$
(S15)

for all the integration domain of equation (S6), so $D_{j_0,x}(k) = 0$, and from equation (S10a), $A = 0$. As one is not able to attribute an argument to the complex number $A$, $\Delta = \arg(AB^*) = \theta_A - \theta_B$ becomes indefinite for a true BIC in the set. As a consequence, from equations (S13a-c), the Fano asymmetry parameter $q$ is also ill-defined, where no value can be attributed to it[1]. However, for an odd mode, the contribution of pairs of in-plane inverted coordinates in equation (S15) only cancel out if a symmetry in the permittivity also exists, i.e.:

$$\varepsilon(k,x',y',z') = \varepsilon(k,-x',-y',z') \quad (S16)$$

Which means that the true BIC cannot be sustained in a system of asymmetric permittivity relative to an in-plane inversion transformation. Or, that it only exists in the limit:

$$\varepsilon(k,x',y',z') \to \varepsilon(k,-x',-y',z') \quad (S17)$$

where $D_{j_0,x}(k)$ becomes zero, with $A = 0$ and all the aforementioned consequences for the Fano parameter $q$.

## 2 – Quality factor versus asymmetry of $\varepsilon$-qBICs

Here we derive rigorously the radiative quality factor as a function of the permittivity asymmetry for the metasurface considered in Figure S1. It follows from the normalization of the perturbative *quasi*-BIC resonant state[1,2] that its inverse radiation lifetime $\gamma$ can be written as a function of the coupling amplitudes $D_i$ of the in-plane components of the electric field of such state ($E_{rs}$):

$$\gamma/c = |D_x|^2 + |D_y|^2 \quad (S18)$$

where

$$D_i = -\frac{\omega_0}{\sqrt{2S_0}c}\int dr'[\varepsilon(\omega_0,r') - 1]E_{rs,i}(r')e^{ik_0z'}, \quad i = x,y \quad (S19)$$

which, by performig a Taylor expansion of $e^{ik_0z'}$, may be written as a function of the components of momenta $\boldsymbol{p}$ (electric dipole), $\boldsymbol{m}$ (magnetic dipole) and $\widehat{\boldsymbol{Q}}$ (electric quadrupole) of the metasurface:

$$D_x = -\frac{k_0}{\sqrt{2S_0}}\left[p_x - \frac{m_y}{c} + \frac{ik_0}{6}Q_{zx}\right] \quad (S20a)$$

$$D_y = -\frac{k_0}{\sqrt{2S_0}}\left[p_y + \frac{m_x}{c} + \frac{ik_0}{6}Q_{yz}\right] \quad (S20b)$$

For the unit cell shown in Figure S1, which is symmetric with respect to the $(x) \to (-x)$ transformation, the resonant state in-plane field components $E_{rs,x}$ and $E_{rs,y}$ are even and odd functions relative to this symmetry, respectively (Figure S2). Being $E_{rs,y}$ an odd function, the contribution of pairs of coordinates $(x, y)$ and $(-x, y)$ to the integral in equation (S19) cancel each other out, thus $D_y = 0$. This argument holds regardless on how much the symmetry in the permittivity is broken, given that the asymmetry in our case demands an in-plane inverse transformation $(x, y) \to (-x, -y)$. Lastly, the even up-down parity of $E_{rs,x}$ ($E_{rs,x}(z) = E_{rs,x}(-z)$) implies that $m_y$ and $Q_{zx}$ are equal to zero:

$$m_y = -\frac{i\omega_0}{2}\int d\mathbf{r}'[\varepsilon(\omega_0, \mathbf{r}') - 1]E_{rs,x}(\mathbf{r}')z' = 0 \quad (S21a)$$

$$Q_{zx} = 3\int d\mathbf{r}'[\varepsilon(\omega_0, \mathbf{r}') - 1]E_{rs,x}(\mathbf{r}')z' = 0 \quad (S21b)$$

given that the integrand is odd and the contributions of coordinate pairs $(x, y, z)$ and $(x, y, -z)$ also sum to zero. This conclusion is also applicable to our case, as the permittivity of the unit cell is symmetric relative to a $(z) \to (-z)$ transformation. It results for equation (S20a):

$$D_x = -\frac{k_0}{\sqrt{2S_0}}[p_x] \quad (S22)$$

then, for equation (S18):

$$\frac{\gamma}{c} = \frac{\gamma_{rad}}{c} = |D_x|^2 = \frac{k_0^2}{2S_0}|p_x|^2 \quad (S23)$$

Here $\gamma_{rad}$ emphasizes that losses are due to radiative processes only. For an isotropic variation in the permittivity $\Delta\varepsilon$, the net dipole moment $p_x$ of the two bars in the unit cell may be calculated as:

$$p_x = \int d\mathbf{r}'[\varepsilon(\omega_0, \mathbf{r}') - 1]E_{rs,x}(\mathbf{r}') \quad (S24)$$

which may be split into two terms, corresponding to each meta-atom:

$$p_x = \int_{meta-atom\ \#1} d\mathbf{r}'[\varepsilon(\omega_0, \mathbf{r}') - 1]E_{rs,x}(\mathbf{r}')$$
$$+ \int_{meta-atom\ \#2} d\mathbf{r}'[\varepsilon(\omega_0, \mathbf{r}') + \Delta\varepsilon - 1]E_{rs,x}(\mathbf{r}') \quad (S25)$$

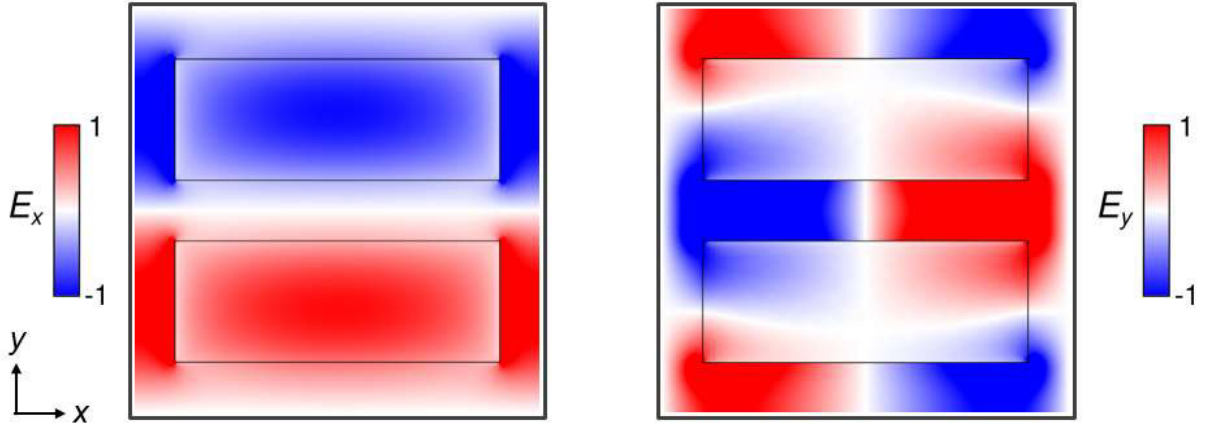

Figure S2: Electric-field components $E_{rs,x}$ (left) and $E_{rs,y}$ (right) of the $\varepsilon$-qBIC.

then

$$p_x = \int_{meta-atom\ \#1} d\mathbf{r}'[\varepsilon(\omega_0, \mathbf{r}') - 1]E_{rs,x}(\mathbf{r}')$$
$$+ \int_{meta-atom\ \#2} d\mathbf{r}'[\varepsilon(\omega_0, \mathbf{r}') - 1]E_{rs,x}(\mathbf{r}') \quad (S26)$$
$$+ \int_{meta-atom\ \#2} d\mathbf{r}'[\Delta\varepsilon]E_{rs,x}(\mathbf{r}')$$

where the first two integrals sum to zero, given the odd parity of $E_{rs,x}$ relative to the $(y) \rightarrow (-y)$ transformation, implying a zero net dipole moment of the unit cell in the absence of the permittivity asymmetry (true BIC condition). The net dipole moment of the unit cell is then:

$$p_x = \int_{meta-atom\ \#2} d\mathbf{r}'[\Delta\varepsilon]E_{rs,x}(\mathbf{r}') \quad (S27)$$

From equation (S23):

$$\frac{\gamma_{rad}}{c} = \frac{k_0^2}{2S_0} \left| \int_{meta-atom\ \#2} d\mathbf{r}'[\Delta\varepsilon]E_{rs,x}(\mathbf{r}') \right|^2 \quad (S28)$$

It results, finally, for the radiative quality factor ($Q_{rad}$):

$$Q_{rad} = \frac{\omega_0}{\gamma_{rad}} = \frac{2S_0}{k_0} \left| \int_{meta-atom\ \#2} d\mathbf{r}'[\Delta\varepsilon]E_{rs,x}(\mathbf{r}') \right|^{-2} \quad (S29)$$

Being the integral independent of $\Delta\varepsilon$:

$$Q_{rad}(\Delta\varepsilon) = \frac{2S_0}{k_0} \left| \int_{meta-atom\ \#2} d\mathbf{r}' E_{rs,x}(\mathbf{r}') \right|^{-2} |\Delta\varepsilon|^{-2} \quad (S30)$$

where an inverse-square dependence of the $Q_{rad}$ with the difference in permittivity between the resonators is obtained (equation 2 of the main manuscript). This dependence is identical to the seemingly universal behavior of geometrically-asymmetric unit cells as a function of their respective asymmetry parameters[1].

### 3- The equivalence of permittivity- and geometrically-asymmetric systems

Given that the different qBICs stem from the same true BIC limiting case, it is reasonable to enquire about the extension of their equivalence. To compare the results derived for the permittivity-asymmetric case with a geometrically-asymmetric system, let us consider a metasurface containing two bars in the unit cell of identical permittivity $\varepsilon(\omega, r)$ whose size vary by an amount $\Delta L$. Considering the symmetry of the system relative to the $(x) \to (-x)$ transformation and the odd parity of the $y$-component of the resonant state electric field $E_{rs,y}$ relative to this symmetry, the net dipole moment of the system $p_x$ can be calculated as in equation (S24):

$$p_x = \int d\mathbf{r}'[\varepsilon(\omega_0, \mathbf{r}') - 1]E_{rs,x}(\mathbf{r}') \tag{S31}$$

An integral which may be split into the domains:

$$p_x = \int_{meta-atom\ \#1} d\mathbf{r}'[\varepsilon(\omega_0, \mathbf{r}') - 1]E_{rs,x}(\mathbf{r}')$$
$$+ \int_{meta-atom\ \#2} d\mathbf{r}'[\varepsilon(\omega_0, \mathbf{r}') - 1]E_{rs,x}(\mathbf{r}') \tag{S32}$$
$$+ \int_{\Delta L} d\mathbf{r}'[\varepsilon(\omega_0, \mathbf{r}') - 1]E_{rs,x}(\mathbf{r}')$$

The first two integrals sum to zero, given the odd symmetry of $E_{rs,x}$ relative to the $(y) \to (-y)$ transformation, where the contribution of coordinate pairs $(x, y)$ and $(x, -y)$ cancel out, corresponding to the true BIC condition, as previously discussed. Thus, the net dipole becomes:

$$p_x = \int_{\Delta L} d\mathbf{r}'[\varepsilon(\omega_0, \mathbf{r}') - 1]E_{rs,x}(\mathbf{r}') \tag{S33}$$

Being the permittivity $\varepsilon(\omega_0, \mathbf{r}') = \varepsilon(\omega_0)$ isotropic in the integration domain:

$$p_x = [\varepsilon(\omega_0) - 1] \int_{\Delta L} d\mathbf{r}' E_{rs,x}(\mathbf{r}') \tag{S34}$$

From equation (S23):

$$\frac{\gamma_{rad}}{c} = \frac{k_0^2}{2S_0} |[\varepsilon(\omega_0) - 1]|^2 \left| \int_{\Delta L} d\mathbf{r}' E_{rs,x}(\mathbf{r}') \right|^2 \tag{S35}$$

Then, for the radiative quality factor:

$$Q_{rad} = \frac{2S_0}{k_0} |[\varepsilon(\omega_0) - 1]|^{-2} \left|\int_{\Delta L} d\mathbf{r}' E_{rs,x}(\mathbf{r}')\right|^{-2} \quad (S36)$$

which has a similar structure to the $Q_{rad}$ derived for a permittivity-asymmetric system (equation (S30)), even though the integrations are performed in different domains. Here, the asymmetry parameter is embedded in the rightmost term. As an approximation, we may consider $E_{rs,x}$ constant within the asymmetry region $\Delta L$, becoming the integration independent of the electric field. The smaller the geometrical asymmetry, the better this approximation holds. Thus:

$$\begin{aligned}Q_{rad} &= \frac{2S_0}{k_0} |[\varepsilon(\omega_0) - 1]|^{-2} |E_{rs,x}|^{-2} \left|\int_{\Delta L} d\mathbf{r}'\right|^{-2} \\ &= \frac{2S_0}{k_0} |[\varepsilon(\omega_0) - 1]|^{-2} |E_{rs,x}|^{-2} |V_{\Delta L}|^{-2}\end{aligned} \quad (S37)$$

where $V_{\Delta L}$ is the volume of the $\Delta L$ domain, that can be written as $V_{\Delta L} = \sigma \Delta L$, being $\sigma$ the lateral cross-section of each meta-atom. It results for $Q_{rad}$:

$$Q_{rad}(\Delta L) = \frac{2S_0}{k_0} |[\varepsilon(\omega_0) - 1]|^{-2} |E_{rs,x}|^{-2} |\sigma|^{-2} |\Delta L|^{-2} \quad (S38)$$

from which a $1/\Delta L^2$ dependence is obtained, as previously observed[1]. Naturally, from it one may define a dimensionless asymmetry parameter $\alpha_L = \Delta L/L_0$ by multiplying both sides of the equation by the square of the original length of the bars $L_0$. We may use the same approach for $Q_{rad}$ of the permittivity-asymmetric system. Using a coarser approximation that the field $E_{rs,x}(\mathbf{r}') = E_{rs,x}$ is constant within the whole meta-atom #2, equation (S30) becomes:

$$\begin{aligned}Q_{rad}(\Delta \varepsilon) &= \frac{2S_0}{k_0} \left|\int_{meta-atom\ \#2} d\mathbf{r}' E_{rs,x}(\mathbf{r}')\right|^{-2} |\Delta \varepsilon|^{-2} \\ &= \frac{2S_0}{k_0} |E_{rs,x}|^{-2} \left|\int_{meta-atom\ \#2} d\mathbf{r}'\right|^{-2} |\Delta \varepsilon|^{-2} \\ &= \frac{2S_0}{k_0} |E_{rs,x}|^{-2} |\sigma|^{-2} |L_0|^{-2} |\Delta \varepsilon|^{-2}\end{aligned} \quad (S39)$$

in which the integration results in the volume of the meta-atom #2, $\sigma L_0$. By equating both radiative quality factors (equations (S38) and (S39)), we have finally:

$$\frac{2S_0}{k_0} |E_{rs,x}|^{-2} |\sigma|^{-2} |L_0|^{-2} |\Delta \varepsilon|^{-2} = \frac{2S_0}{k_0} |[\varepsilon(\omega_0) - 1]|^{-2} |E_{rs,x}|^{-2} |\sigma|^{-2} |\Delta L|^{-2} \quad (S40)$$

that might be rearranged into:

$$\frac{2S_0}{k_0}|E_{rs,x}|^{-2}|\sigma|^{-2}\left|\frac{\Delta\varepsilon}{[\varepsilon(\omega_0)-1]}\right|^{-2} = \frac{2S_0}{k_0}|E_{rs,x}|^{-2}|\sigma|^{-2}\left|\frac{\Delta L}{L_0}\right|^{-2}$$

$$\frac{2S_0}{k_0}|E_{rs,x}|^{-2}|\sigma|^{-2}|\alpha'_\varepsilon|^{-2} = \frac{2S_0}{k_0}|E_{rs,x}|^{-2}|\sigma|^{-2}|\alpha_L|^{-2}$$

(S41)

which are identical as long as the redefined asymmetry parameter $\alpha'_\varepsilon$ and $\alpha_L$ have the same value, as confirmed by numerical analysis (see Figure 2d of the main manuscript). The equivalence between the asymmetric systems holds the better is the approximation of the constant electric field $E_{rs,x}$ within the meta-atoms. Note that we may define a common coefficient $Q_{0_{\varepsilon,L}}$:

$$Q_{0_{\varepsilon,L}} = \frac{2S_0}{k_0}|E_{rs,x}|^{-2}|\sigma|^{-2}$$

(S42)

implying that both qBICs stem from the same symmetry-protected BIC.

### 4 – Ellipsometry data of Si and TiO$_2$ thin films

Figure S3 shows the complex permittivity obtained from the fitted data of ellipsometry measurements of plasma-enhanced chemical vapor deposition (PECVD)-grown Si and TiO$_2$ thin films. These were used in the subsequent etching procedures for the fabrication of the resonators of the metasurfaces used in experiments.

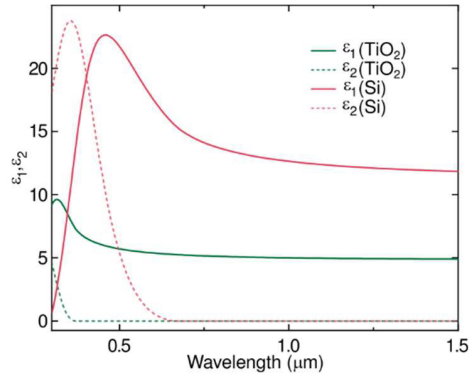

Fig S3: Real ($\varepsilon_1$) and imaginary ($\varepsilon_2$) parts of the permittivity of Si (red) and TiO$_2$ (green) films used in samples fabrication.

# 5 – Fano parameter $q$ of a geometric asymmetry compensation via permittivity symmetry breaking

Figure S4 shows the values of the Fano parameter $q$ as a function of the permittivity asymmetry $\alpha_\varepsilon$ for the resonances displayed at the bottom panel of Fig. 4e of the main manuscript. The compensation of an arbitrary $10\ nm$ difference in height between ridges in the unit cell of a 1DG using asymmetries in the permittivity results in $q$ changing sign twice (from negative to positive, and back to negative again) for the calculated range.

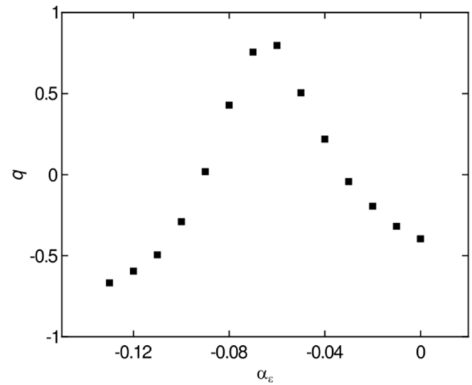

Fig. S4: Fano parameter $q$ as a function of the permittivity asymmetry $\alpha_\varepsilon$.